\documentclass[aps,reprint,prd,a4paper,superscriptaddress,nofootinbib,preprintnumbers]{revtex4-1}
\usepackage[utf8]{inputenc}
\usepackage{amsmath}
\usepackage{graphicx}
\usepackage{aas_macros}

\begin{document}

\title{Explaining cosmic ray antimatter with secondaries from old supernova remnants}

\author{Philipp Mertsch}
\affiliation{Institute for Theoretical Particle Physics and Cosmology (TTK), RWTH Aachen University, 52056 Aachen, Germany}

\author{Andrea Vittino}
\affiliation{Institute for Theoretical Particle Physics and Cosmology (TTK), RWTH Aachen University, 52056 Aachen, Germany}

\author{Subir Sarkar}
\affiliation{Rudolf Peierls Centre for Theoretical Physics, University
  of Oxford, Parks Road, Oxford OX1 3PU, UK} 

\preprint{TTK-20-51}

\begin{abstract}
Despite significant efforts over the past decade, the origin of the cosmic ray positron excess has still not been unambiguously established. A popular class of candidate sources are pulsars or pulsar wind nebulae but these cannot also account for the observed hard spectrum of cosmic ray antiprotons. We revisit the alternative possibility that the observed high-energy positrons are secondaries created by spallation in supernova remnants during the diffusive shock acceleration of the primary cosmic rays, which are further accelerated by the same shocks. The resulting source spectrum of positrons at high energies is then naturally harder than that of the primaries, as is the spectrum of other secondaries such as antiprotons. We present the first comprehensive investigation of the full parameter space of this model --- both the source parameters as well as those governing galactic transport. Various parameterisations of the cross sections for the production of positrons and antiprotons are considered, and the uncertainty in the model parameters discussed. We obtain an  excellent fit to recent precision measurements by AMS-02 of cosmic ray protons, helium, positrons and antiprotons, as well as of various primary and secondary nuclei. This model thus provides an economical explanation of the spectra of \emph{all} secondary species --- from a single well-motivated population of sources.
\end{abstract}


\maketitle

\section{Introduction}

It is generally believed~\cite{Gabici:2019jvz} that strong shocks in supernova remnants (SNRs) accelerate the bulk of Galactic cosmic rays (CRs) by the diffusive shock acceleration (DSA) mechanism, out of ambient matter that has close to solar elemental abundances. Abundant elements like H, He, C and O thus constitute the CR `primaries', while Be, B and Li as well as some sub-iron elements which are expected to be absent in the sources and can be created only via spallation are termed `secondaries'. N is known to receive both primary and secondary contributions. Positrons and antiprotons observed in CR are also taken to be secondaries. The transport of CRs from the sources through the Galaxy is then probed by measuring the relative fraction of secondaries produced by the spallation of primaries on gas in the interstellar medium (ISM). The mass density integrated along CR trajectories is the `grammage' which is expected to decrease with increasing rigidity $\mathcal{R}$ ($\equiv p c / Z e$), reflecting the diffusive nature of CR transport, with a diffusion coefficient that grows, typically as a power-law, with rigidity.

Traditionally, the grammage traversed by CRs \emph{in} the sources is neglected since the SNR lifetime, which bounds the time spent by CRs in the source, is  much smaller than their typical propagation time in the Galaxy. Even with a higher gas density near the shocks, (\textit{e.g.}, due to interaction with molecular clouds), the source grammage remains smaller than the grammage accumulated in the ISM. What this argument does not take into account however is that CR secondaries produced close to the shock will necessarily also undergo further acceleration by the \emph{same} shock that accelerates the primaries. The fraction of secondaries which partake are essentially those within a diffusion length of the shock. The effective source term for secondaries produced during DSA thus has a rigidity-dependence which follows that of the diffusion coefficient near the shock.

In recent years several observational anomalies have emerged concerning CRs, the most prominent being the `positron excess' first identified by the PAMELA satellite~\cite{Adriani:2008zr,Aguilar:2013qda}. As mentioned, the positron fraction, {\it i.e.} the ratio of secondary positrons to the primary electrons (plus positrons), should fall with increasing rigidity, reflecting the smaller grammage traversed in the Galaxy at higher rigidity. Instead the positron fraction was seen to rise at high rigidity, most recently by the AMS-02 experiment on the International Space Station~\cite{Aguilar:2019owu}, thus implicating a new source of positrons with a harder spectrum. This has been widely attributed to new physics beyond the Standard Model, \textit{viz.} the annihilation or decay of dark matter constituted of relic electroweak scale mass particles. An alternative astrophysical explanation invokes pulsars (or rather pulsar wind nebulae, PWNe) as the source of additional \emph{primary}  positrons~\cite{Hooper:2008kg,Yuksel:2008rf,Profumo:2008ms,Grasso:2009ma,Evoli:2020szd,DiMauro:2020cbn}. The observation by HAWC and Fermi-LAT of a $\gamma$-ray halo around Geminga does demonstrate some contribution by PWNe to the observed positron flux \cite{Linden:2017vvb,Abeysekara:2017old,DiMauro:2019yvh}, however their overall contribution is hard to estimate reliably. Positrons with the required hard spectrum are however \emph{naturally} produced by CR spallation and subsequent acceleration in nearby SNRs~\cite{Berezhko:2003pf,Blasi:2009hv,Mertsch:2009ph,Ahlers:2009ae,Tomassetti:2012ir,Mertsch:2014poa,Tomassetti:2017izg} as we reexamine in this paper.

Concerning CR antiprotons, there has been much speculation about an excess around $10\,\text{GeV}$ with explanations ranging from correlated systematic errors to dark matter annihilation. Comparatively less attention has been paid to another anomaly \textit{viz.} the unexpected hardness of the antiproton spectrum. By the same arguments as for positrons or secondary nuclei, the (secondary) antiproton spectrum ought to be softer than the primary proton spectrum. However, measurements by AMS-02 up to $\sim500\,\text{GeV}$ show the antiproton spectrum to be just as hard as that of protons. While this can be explained by assuming the diffusion coefficient in interstellar space to have a weak rigidity dependence, this is unexpected and was not discussed before the AMS-02 measurement~\cite{Lipari:2018usj}. See, however, \cite{Boudaud:2019efq}.) Again the acceleration of secondary antiprotons in SNRs can naturally account for the observed hard spectrum.

The model of acceleration of secondaries in SNRs is especially economical in that no new population of sources needs to be invoked to explain the positron excess. It is also very predictive in that the acceleration should affect all secondaries \emph{equally}. A hard positron spectrum is therefore necessarily accompanied by a hard spectrum of antiprotons, as well as of Li, Be and B and  sub-Fe elements. Previous studies of this model have had difficulties in accommodating all the data, however the adopted parameters to describe CR propagation in the Galaxy had been obtained under the assumption that there is \emph{no} acceleration of secondaries~\cite{Cholis:2013lwa}. It is clear that for a self-consistent analysis it is essential to \emph{simultaneously} fit both the source and propagation parameters.

Despite its general predictive nature, the `acceleration of secondaries' model has a large number of free parameters which need to be determined by fitting to data. Whereas conventional models of CR typically have ten or so parameters (e.g.~\cite{Weinrich:2020cmw,Heisig:2020nse,Boschini:2020jty}), it is now necessary to add several more. This is because of  uncertainties in modelling for instance the diffusion rate in the sources. Other uncertainties are related to the production cross section of electrons, positrons, antiprotons and secondary nuclei, although there has been significant recent progress here in accelerator measurements (e.g.~\cite{Winkler:2017xor}). In any case, navigating the higher-dimensional parameter space becomes increasingly difficult numerically and previous studies may have missed viable regions of parameter space. In this paper we revisit the issue and find parameters that consistently reproduce the observations by AMS-02 of p and He, B and C, as well as of positrons and antiprotons. 

This paper is organised as follows: in \S~\ref{sec:accsec_theory} we recapitulate the essential theory and discuss the cross section models employed as well as propagation parameters. Then in \S~\ref{sec:Analysis} we fit to the recent precision data from AMS-02, as well as the interstellar spectra measured by Voyager~1, and discuss the goodness of fit. In \S~\ref{sec:Summary} we present our summary and conclusions.

\section{Method}
\label{sec:accsec_theory}

\subsection{Acceleration of cosmic rays in the sources}

Strong SNR shocks are believed to accelerate  galactic CRs up to the `knee' in their energy spectrum at $\sim 3 \times 10^{6}\,\mathrm{GeV}$. While being accelerated by the DSA mechanism, CR particles undergo collisions with the ambient gas resulting in the production of secondaries. The standard lore neglects this contribution on the grounds that the grammage  traversed within the source is very small compared to the grammage while diffusing in the ISM (e.g.~\cite{Katz:2009yd}). However the secondaries produced in the source, although subdominant in number, can have a different spectrum than secondaries produced during propagation, and thus make an important contribution in a specific energy range. In particular, the secondaries created in the source are \emph{necessarily} accelerated by the same shock that accelerates the primaries. However while primaries are injected only at the SNR shock, the secondaries that participate in DSA are present in a wider region, approximately up to a diffusion length away from the shock. Because of this, secondaries that undergo acceleration in the sources acquire a \emph{harder} spectrum and thus come to dominate at high energies. We recapitulate below the key steps in quantifying this contribution. 

Following our previous work \cite{Mertsch:2009ph,Ahlers:2009ae,Mertsch:2014poa} we adopt the DSA test-particle approximation. We choose to work in  the rest-frame of the shock, which is located at $x = 0$, with $x < 0$ being upstream and $x>0$ downstream. The compression ratio of the shock, fixed by the Rankine-Hugoniot conditions, is $r = n_+ / n_- = u_- / u_+$ with $n_{\pm}$ and $u_{\pm}$ being the gas densities and velocities downstream and upstream of the shock. The evolution of the phase space density for CR species $i$, averaged over the gyro-phase and pitch-angle, is then governed by the equation:
\begin{equation}
    \frac{\partial f_i}{\partial t} = -u \frac{\partial f_i}{\partial x} + \frac{\partial}{\partial x} D_i \frac{\partial f_i}{\partial x} - \frac{p}{3} \frac{du}{dx} \frac{\partial f_i}{\partial p} - \Gamma_i f_i + q_i.
\label{eq:CR_transport}
\end{equation}
The terms on the rhs describe, respectively, convection, spatial diffusion, adiabatic momentum losses/gains, inelastic losses and injection. If the species $i$ is of secondary origin, the injection term $q_i$ takes into account spallation processes that involve heavier species by summing over all such contributions: $q_i = c n_{\mathrm{gas}} \sum_{j>i} \beta_j \sigma_{i\rightarrow j}$. We follow Ref.~\cite{Herms:2016vop} in not assuming a fixed inelasticity, but using the distribution implied by the differential cross sections (see also Ref.~\cite{Kachelriess:2011qv}).

Eq.~(\ref{eq:CR_transport}) may be solved in the steady-state regime, both upstream and downstream of the shock \cite{Mertsch:2009ph,Ahlers:2009ae,Mertsch:2014poa}. Under the conditions $\Gamma_i D_i/u_\pm^2 \ll 1$ and $x \Gamma_i/u_{\pm} < x_{\mathrm{max}} \Gamma_i/u_{\pm} \ll 1$, the  solution downstream is:
\begin{equation}
f_i^+(x,p) = f_i^0(p) + r\left( q_i^0(p) - \Gamma_i^- f_i^0(p)\right) \frac{x}{u_+}
\end{equation}
where $q_i^0(p)$ and $f_i^0(p)$ denote, respectively, the injection term and the phase space density at the shock, with: 
\begin{equation}
\begin{aligned}
    f_i^0(p) = \int_0^p \frac{dp'}{p'} \left( \frac{p'}{p}\right)^{\gamma}\,e^{ -\gamma(1+r^2) \left( D_i(p) - D_i(p')\right)\frac{\Gamma_i(p)}{u^2_-}} &\\
    \times \left[ \gamma (1 + r^2) \frac{D_i(p')}{u^2_-}q_i^0(p') + \gamma Y_i \delta(p'-p_0) \right]. &
\end{aligned}
\end{equation}
If we assume that all particles are released in the ISM after a time $\tau_{\mathrm{SNR}}$, which is the effective lifetime of the SNR, the integrated spectrum downstream is:
\begin{equation}
    \begin{aligned}
    \frac{dN_i}{dp} &= 4 \pi\int_0^{\tau_{\mathrm{SNR}}u_+} dx x^2 4 \pi p^2 f_i(x,p)\\
                &= 4\pi V^2 \left[ f_i^0 + \frac{3}{4} r \tau_{\mathrm{SNR}}\left(q_i^0 - \Gamma_i^- f_i^0 \right) \right]
    \end{aligned}
\label{eq:downstream_spectrum}
\end{equation}
with $V = 4/3 \pi (u_{+}\tau_{\mathrm{SNR}})^3$ being the downstream volume.

An important parameter is $\mathcal{R}_{\mathrm{max}}$, the maximum rigidity to which CRs can be accelerated by the SNR. This is determined by either the diffusion coefficient and the shock velocity, or by the SNR age, according to whether the DSA scenario is `age-limited' or `escape-limited'. Following earlier reasoning \cite{Mertsch:2014poa}, we take $\mathcal{R}_{\mathrm{max}}$ to be in the range 1--10~TeV. We assume that the SNR density $f(r)$ varies with galacto-centric radius $r$ as: 
\begin{equation}
f(r) \propto \left( \frac{r}{R_{\odot}} \right)^{0.48} \exp \left[ -2.2 \left( \frac{r}{R_{\odot}} - 1 \right) \right] \, ,
\end{equation}
for $r \leq 10 \, \text{kpc}$ (where  $R_{\odot} = 8.5 \, \text{kpc}$). For $10 \, \text{kpc} < r \leq 15 \, \text{kpc}$, $f = f(10 \, \text{kpc})$ and $f=0$ beyond~\cite{Strong:2010pr}.

\subsection{Production of antiprotons and positrons}

\begin{figure*}[ht]
    \centering
    \includegraphics[width=0.45\textwidth]{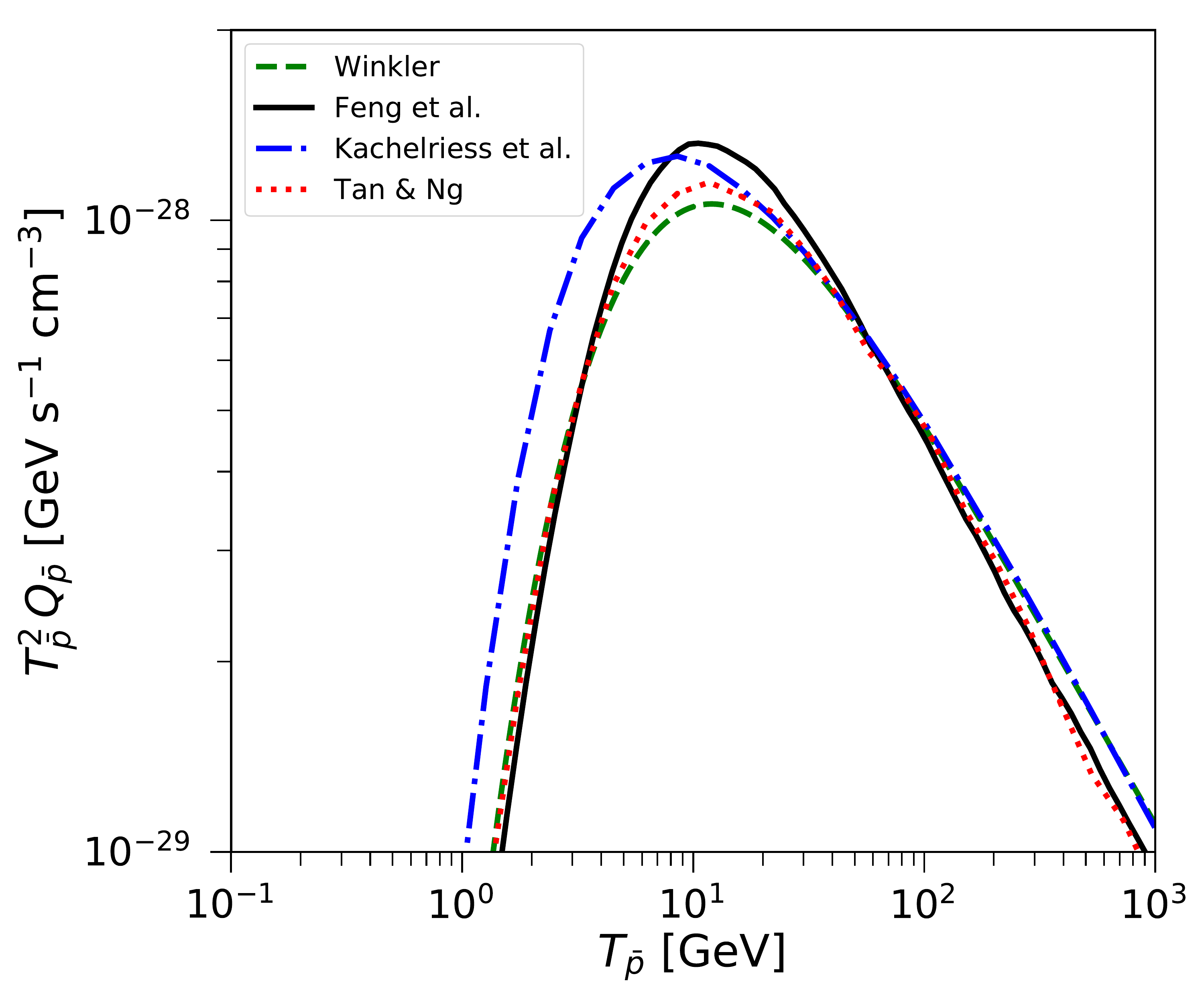}
    \includegraphics[width=0.45\textwidth]{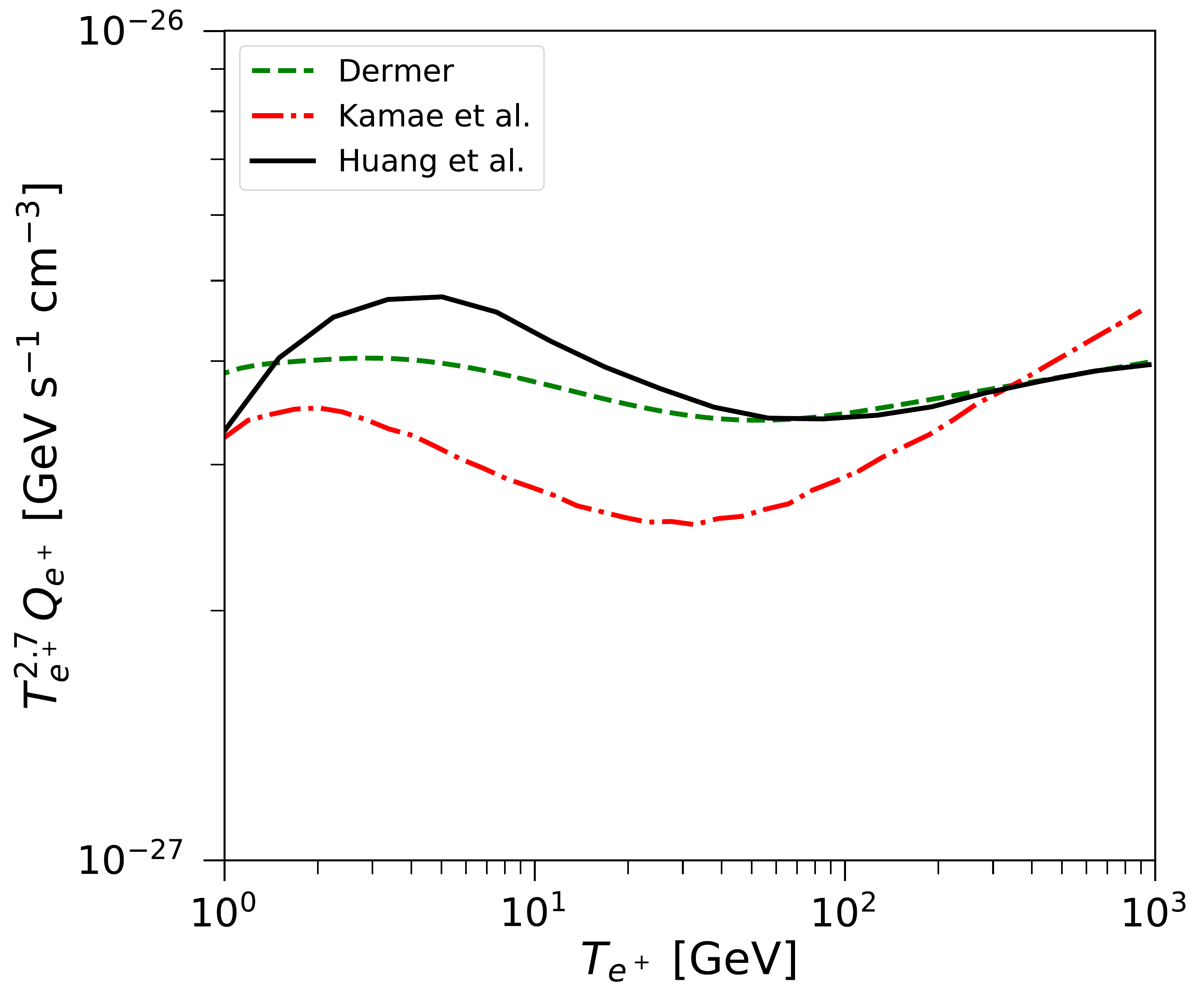}
    \caption{Source terms for the production of secondary CR antiprotons ({\it left panel}) and positrons ({\it right panel}) as predicted by different models. For antiprotons we show the source term associated to the pp channel only, while for positrons we show the total source term, with the helium contribution for the Dermer and Kamae \textit{et al.} models computed by adopting the rescaling proposed  \cite{Norbury:2006hp} for the Huang \textit{et al.} model  \cite{Huang:2006bp}. From~\cite{Evoli:2017vim}.}
    \label{fig:source_terms}
\end{figure*}

The modelling of the  production of secondary antiprotons and positrons plays a central role in our fit to the data. We find that the various proposed prescriptions in the literature yield very different outcomes for the final CR spectra, although they are all tuned to fit the same accelerator data.  

Typically two categories of models are in use: those based on semi-empirical parameterisations, and those based on Monte Carlo event generators. For the former, the cross section is described in terms of a parametric formula inspired by some underlying physical principle ({\it e.g.},  Feynman scaling) and the parameters are tuned to reproduce the available experimental data. For the latter, the cross section is obtained by simulating with a Monte Carlo (MC) event generator the process under study, \textit{viz.} fixed-target collisions with either proton or helium as projectile/target in the present case. The many parameters of the MC event generator are tuned to reproduce existing experimental data.

We use the reference model for secondary antiproton production \cite{Feng:2016loc} that uses the \texttt{EPOS LHC} event generator \cite{Pierog:2013ria}, which is in best agreement with data from the NA49 \cite{Fischer:2003xh}, BRAHMS \cite{Arsene:2007jd} and ALICE \cite{Aamodt:2011zj} experiments. In this model, all 4 reactions which contribute to the production of secondary antiprotons (\textit{viz.} pp, pHe, Hep and HeHe collisions) are  obtained directly from \texttt{EPOS LHC}. 

We have compared the results of our reference model to three others --- two of which are semi-empirical parametrisations (by Tan and Ng \cite{Tan:1982nc} and by Winkler \cite{Winkler:2017xor}), and one based on \texttt{QGSJET-II} \cite{Kachelriess:2015wpa}. The corresponding secondary source terms associated with pp spallation reactions are shown in the left panel of Fig.~\ref{fig:source_terms}. Of the two semi-empirical parametrisations, the model \cite{Tan:1982nc} assumes strict radial scaling while the model \cite{Winkler:2017xor} allows for the violation of such scaling. Thus, as seen in the figure, the two parametrisations give rather different results at high energies where the scaling violation occurs. In addition, the model \cite{Winkler:2017xor} takes into account isospin violation which causes pp collisions to produce more p$\bar{\mathrm{n}}$ pairs than n$\bar{\mathrm{p}}$ --- generating an asymmetry between the production of secondary antiprotons and secondary antineutrons. That the effect indeed exists can be inferred by comparing \cite{Fischer:2003xh} the proton yield of n + p collisions with the neutron yield of p + p collisions in the NA49 experiment. As discussed elsewhere \cite{Feng:2016loc}, while this effect is in general not present in MC event generators, our reference model which makes use of \texttt{EPOS LHC} predicts an asymmetric production of $\bar{n}$ and $\bar{p}$ with the ratio between the cross sections $\sigma_{pp \rightarrow \bar{n}}$ and $\sigma_{pp \rightarrow \bar{p}}$ ranging between 1 and 1.9 depending on the energy of the process. Finally the model \cite{Kachelriess:2015wpa} is based on a specific tune of \texttt{QGSJET-II}, labelled \texttt{QGSJET-II-m}, which has been shown to provide a more accurate description of low-energy processes. The source term given by this model \cite{Kachelriess:2015wpa} differs significantly at low energies from the others under study, as seen in Fig.~\ref{fig:source_terms}; at high energies however it closely follows the model \cite{Winkler:2017xor}.   

For secondary positrons too one can choose between semi-empirical parameterisations and models based on MC event generators. Secondary positrons are produced through the decay of charged or neutral pions which, in turn, are produced either directly in the collision or as the result of the decay of other particles. The modelling of the pion production cross section is conceptually similar to the modelling of the antiproton production cross section and is thus described similarly. Here, semi-empirical parameterisations are typically based on the scaling hypothesis \cite{Badhwar:1977zf,Stephens:1981fr,1983JPhG....9.1289T}. Among these parameterisations, the one most used \cite{1986A&A...157..223D,1986ApJ...307...47D} combines the model \cite{Badhwar:1977zf,Stephens:1981fr} with the `isobaric' model \cite{1969ApJ...157..507S} at low-energies. This is the default choice in the \texttt{GALPROP} code. A recent refinement of this parameterisation for the production of neutral pions at high transverse momentum has been proposed \cite{,Blattnig:2000zf}. In an effort to overcome some potential  shortcomings of the semi-empirical parametrisations, MC event generators are often used. This is the case for the models proposed by Kamae \textit{et al.}  \cite{Kamae:2004xx,Kamae:2006bf} and Huang \textit{et al.} \cite{Huang:2006bp}, the former based on \texttt{PYTHIA 6.2} and the latter on \texttt{DPMJET-III}. We adopt the model \cite{Huang:2006bp} in the present work. The secondary positron source terms obtained with the different prescriptions are shown in the right panel of Fig.~\ref{fig:source_terms}. As can be seen, there is a fairly large difference between the two models based on MC event generators, with our adopted model \cite{Huang:2006bp}  providing the largest yield of positrons over the energy interval from 1 to 300 GeV.

\subsection{Cosmic-ray propagation and solar modulation}

To model CR transport in the ISM we use the standard \texttt{GALPROP} code \cite{Moskalenko:1997gh,Strong:1998pw}, which performs a numerical solution of the CR transport equation. This is the same as Eq.~(\ref{eq:CR_transport}), with the source term being the downstream spectrum defined in Eq.~(\ref{eq:downstream_spectrum}) and all remaining parameters (describing spatial diffusion, energy losses and gas densities) adapted to describe the ISM. 

In our treatment of CR Galactic transport, special attention has to be paid to the spatial diffusion coefficient, which we assume to be characterised by two breaks in rigidity as motivated in the recent model \cite{Vittino:2019yme}: 
\begin{equation}
D_{xx}(\mathcal{R}) \! = \! D_0 \beta \! \left( \! \frac{\mathcal{R}}{\mathcal{R}_1} \! \right)^{\delta_1} \!\! \prod_{i=1}^2 \left( \! 1 + \left( \! \frac{\mathcal{R}}{\mathcal{R}_i} \! \right)^{1/s_i} \! \right)^{\! s_i (\delta_{i+1} - \delta_i)} \! . \!\!
\label{eq:diff_coefficient}
\end{equation}
A physical motivation for the low-rigidity break is the damping of turbulence caused by an almost isotropic cosmic-ray distribution \cite{Ptuskin:2005ax}, while the high-rigidity break corresponds to a change in the source of the turbulence that is responsible for cosmic-ray diffusion \cite{Blasi:2012yr}.

To describe solar modulation of charged CRs entering the heliosphere, we adopt the force-field approximation \cite{Gleeson:1968zza} with different Fisk potentials  for different CR species, as will be illustrated in more detail in the next section.

\section{Analysis}
\label{sec:Analysis}
As discussed in \S~\ref{sec:accsec_theory}, the spectra of secondaries which are accelerated further in the sources depend on several parameters: the shock compression ratio $r = u_{-}/u_{+}$, the SNR age $\tau_{\mathrm{SNR}}$, the gas density $n_{\mathrm{gas}}$ and the diffusion coefficient near the shock:
\begin{equation}
    D = 3 \times 10^{22} \mathrm{cm}^2\, \mathrm{s}^{-1}\, \beta \, K_\mathrm{B} \left(\frac{B}{1 \, \mu\text{G}} \right)^{-1} \left( \frac{\mathcal{R}}{1 \, \text{GV}} \right)^{\alpha} \, . \label{eqn:source_diff_coeff}
\end{equation}
Here, $K_\mathrm{B}$ quantifies deviations from the Bohm diffusion rate at $1 \, \text{GV}$, and $\alpha$ is the (power law) rigidity scaling of the diffusion coefficient. 

As shown in \S~\ref{sec:accsec_theory}, the combination of parameters entering the secondary source term is $K_\mathrm{B}/{u_{-}^2 B}$. We follow earlier work \cite{Mertsch:2014poa} and fix $B = 1\,\mu \mathrm{G}$ and $u_{-} = 5 \times 10^7$ cm s$^{-1}$, leaving $K_\mathrm{B}$ as the only free parameter that determines the normalisation of the production of secondaries.

\subsection{Fit to experimental data}
\label{sec:fit}

\begin{table*}[!th]
\caption{Free parameters and their best-fit values.}
\small
\label{tbl1}
\begin{tabular}{r l l}
\hline\hline
\multicolumn{3}{c}{Source parameters} \\
\hline
$\tau_{\mathrm{SNR}}$ 			& $= \left({2.664}^{+0.225}_{-0.249}\right) \times 10^4$ 	& source age in yr \\
$K_\mathrm{B}$ 				& $= \left( {1.212}^{+0.293}_{-0.282} \right) \times 10^3$	& normalisation of source diffusion coefficient, see eq.~\eqref{eqn:source_diff_coeff} \\
$\mathcal{R}_{\mathrm{max}}$ 		& $= \left( {1.369}^{+0.278}_{-0.186} \right) \times 10^3$ 	& maximum rigidity in GV \\
$\alpha$ 						& $= {0.612}^{+0.025}_{-0.019}$ 					& spectral index of diffusion coefficient \\
$\gamma^p_1$ 				& $= {1.834}^{+0.018}_{-0.022}$ 					& proton spectral indices below ${\mathcal{R}}^p_{\mathrm{br}}$ \\
$\gamma^p_2$ 				& $= {2.311}^{+0.005}_{-0.006}$ 					& proton spectral indices above ${\mathcal{R}}^p_{\mathrm{br}}$ \\
$\log_{10}\left[ {\mathcal{R}}^p_{\mathrm{br}} / \text{GV} \right]$ 	& $= {3.330}^{+0.022}_{-0.022}$ 					& break rigidity \\
$s_1$						& $= {0.514}^{+0.049}_{-0.047}$					& softness of break at ${\mathcal{R}}^p_{\mathrm{br}}$ \\
$\gamma^{\mathrm{He}}_2$ 		& $= {2.252}^{+0.004}_{-0.005}$ 					& Helium spectral index above ${\mathcal{R}}^{\mathrm{He}}_{\mathrm{br}}$ \\
$\gamma^{\mathrm{C}}_2$ 		& $= {2.259}^{+0.004}_{-0.005}$ 					& Carbon spectral index above ${\mathcal{R}}^{\mathrm{C}}_{\mathrm{br}}$ \\
$\gamma^{\mathrm{nuc}}_2$ 		& $= {2.290}^{+0.004}_{-0.005}$ 					& Nuclear spectral index above ${\mathcal{R}}^{\mathrm{nuc}}_{\mathrm{br}}$ \\
$N$ 							& $= 1.002^{+0.001}_{-0.001}$ 						& Overall scaling factor \\
\hline \multicolumn{3}{c}{Galactic parameters} \\
\hline
$D_0$ 						& $= {4.082}^{+0.109}_{-0.096}$ 					& normalisation of diffusion coefficient in 10$^{28}$ cm$^{2}$s$^{-1}$ \\
$\delta_1$						& $= {-0.250}^{+0.046}_{-0.050}$ 					& spectral index of diffusion coefficient below $\mathcal{R}_{12}$ \\
$\log_{10}\left[ \mathcal{R}_1/ \text{GV} \right]$				& $= {3.586}^{+0.033}_{-0.040}$ 					& first break rigidity \\
$s_1^{\text{diff}}$				& $= {0.251}^{+0.068}_{-0.060}$					& softness of break at $\mathcal{R}_{12}$ \\
$\delta_2$						& $= {0.538}^{+0.007}_{-0.006}$ 					& spectral index of diffusion coefficient between $\mathcal{R}_{12}$ and $\mathcal{R}_{23}$ \\
$s_2^{\text{diff}}$				& $= {10.112}^{+6.863}_{-3.189}$					& softness of break at $\mathcal{R}_{23}$ \\
$v_A$						& $= {25.515}^{+1.669}_{-1.635}$					& Alfv\'en speed in km\,s$^{-1}$\\
\hline \multicolumn{3}{c}{Solar modulation parameters} \\
\hline
$\phi_p$						& $= 0.655^{+0.003}_{-0.003}$ 						& Fisk potential for protons in GV \\	
$\phi_{e^+}$					& $= 0.508^{+0.002}_{-0.002}$ 						& Fisk potential for positrons in GV \\	
$\phi_{\bar{p}}$					& $= 0.425^{+0.019}_{-0.018}$ 						& Fisk potential for antiproton in GV \\	
$\phi_{\mathrm{nuc}}$			& $= 0.664^{+0.003}_{-0.003}$ 						& Fisk potential for nuclei in GV \\	
\hline\hline
\end{tabular}
\end{table*}

To fit experimental data, we perform a Markov Chain Monte Carlo (MCMC) sampling of the parameter space with the \texttt{emcee} package \cite{ForemanMackey:2012ig} which implements the affine-invariant ensemble sampler \cite{2010CAMCS...5...65G}. 

Our parameter space comprises those associated with the SNR environment and with the injected CR spectra, together with those associated with CR transport (in the Galaxy, as well as in the heliosphere). In modelling SNRs, we take to be  free parameters the SNR age $\tau_{\mathrm{SNR}}$, the maximum rigidity $\mathcal{R}_{\mathrm{max}}$ to which (secondary) CRs are accelerated by the SNR, $\alpha$ the rigidity scaling of the spatial diffusion coefficient near the SNR shock, and $K_\mathrm{B}$ the factor by which it is enhanced over the Bohm value at 1 GV. The injected CR spectra are assumed, as is customary, to be (broken) power-laws in rigidity. We also consider as free parameters the slopes $\gamma^p_1$ and  $\gamma^p_2$, as well as the break rigidity ${\mathcal{R}}^p_{\mathrm{br}}$ with $s_1$ parametrising the softness of the break for protons:
\begin{equation}
\frac{\mathrm{d} N}{\mathrm{d}\mathcal{R}} \propto \left( \frac{\mathcal{R}}{\mathcal{R}^p_{\mathrm{br}}} \right)^{\gamma^p_1} \left( 1 + \left( \frac{\mathcal{R}}{\mathcal{R}^p_{\mathrm{br}}} \right)^{s_1} \right)^{(\gamma^p_2 - \gamma^p_1)/s_1} \, .
\end{equation}
While the need for such a spectral break was realised as soon as Voyager data outside the heliosphere became available~\cite{Cummings:2016pdr}, the interpretation of this break in a physical picture of CR acceleration is now all but certain. Source stochasticity effects naturally lead to a break in the propagated spectra~\cite{Phan:2021iht}, which can be emulated by a break in the source spectrum. For the helium and carbon spectrum, we fix $\gamma^{\mathrm{He}}_1 = \gamma^{\mathrm{C}}_1 = \gamma^p_1 - 0.08 $ while letting $\gamma^{\mathrm{He}}_2$ and $\gamma^{\mathrm{C}}_2$ vary. All heavier nuclei are treated in the same way with a free, common high-energy spectral index $\gamma^{\mathrm{nuc}}_2$. The softness of the break is characterised by one parameter, $s_1$. As for the break rigidities of heavier nuclei, we have related these to the one for protons as, $\mathcal{R}^{\mathrm{He}}_{\mathrm{br}}, \mathcal{R}^{\mathrm{C}}_{\mathrm{br}}, \mathcal{R}^{\mathrm{nuc}}_{\mathrm{br}} = 1.7 \, \mathcal{R}^p_{\mathrm{br}}$, which should approximately emulate the effect of source discreteness~\cite{Phan:2021iht} on the propagated fluxes. 

In using Eq.~(\ref{eq:diff_coefficient}) to model diffusion, we take as free parameters the normalisation $D_0$, the slopes at low and intermediate rigidities $\delta_1$, $\delta_2$, and the position of the first rigidity break $\mathcal{R}_1$. We fix $\delta_3 = \delta_2 - 0.17$ and $\mathcal{R}_2 = 300$ GV. In modelling transport in the interstellar medium, we take the Alv\'{e}n speed $v_A$ to be a free parameter and ignore the possible effect of advection. In modelling solar modulation, we adopt different Fisk potentials for different CR species: $\phi_p$, $\phi_{e^+}$, $\phi_{\bar{p}}$ and $\phi_{\mathrm{nuc}}$ denote the potentials used to modulate CR protons, positrons, antiprotons and heavier nuclei (He, B, C, O), respectively. Lastly, we include as a free parameter an overall scaling factor $N$ which is applied to all fluxes. Our model thus has in total 23 free parameters: $\tau_{\mathrm{SNR}}$, $K_\mathrm{B}$, $\mathcal{R}_{\mathrm{max}}$, $\alpha$ , $\gamma^p_1$, $\gamma^p_2$, ${\mathcal{R}}^p_{\mathrm{br}}$, $s_1$, $\gamma^{\mathrm{He}}_2$, $\gamma^{\mathrm{C}}_2$, $\gamma^{\mathrm{nuc}}_2$, $D_0$, $\delta_1$, $\delta_2$, $\mathcal{R}_1$, $s_1^{\text{diff}}$, $s_2^{\text{diff}}$, $v_A$, $\phi_p$, $\phi_{e^+}$, $\phi_{\bar{p}}$, $\phi_{\mathrm{nuc}}$, $N$.

In addition to the precision AMS-02 data on protons \cite{PhysRevLett.114.171103}, helium, carbon and oxygen  \cite{PhysRevLett.119.251101}, antiprotons \cite{PhysRevLett.117.091103}, positrons \cite{PhysRevLett.113.121102}, and boron \cite{PhysRevLett.120.021101} in near-Earth space, we also fit the flux of interstellar CR protons, helium, carbon and boron measured by Voyager~1~\cite{Cummings:2016pdr}. We do not include in the fit the spectra of other CR species measured by Voyager~1.

\subsection{Results}

\begin{figure*}[!thb]
\includegraphics[scale=1]{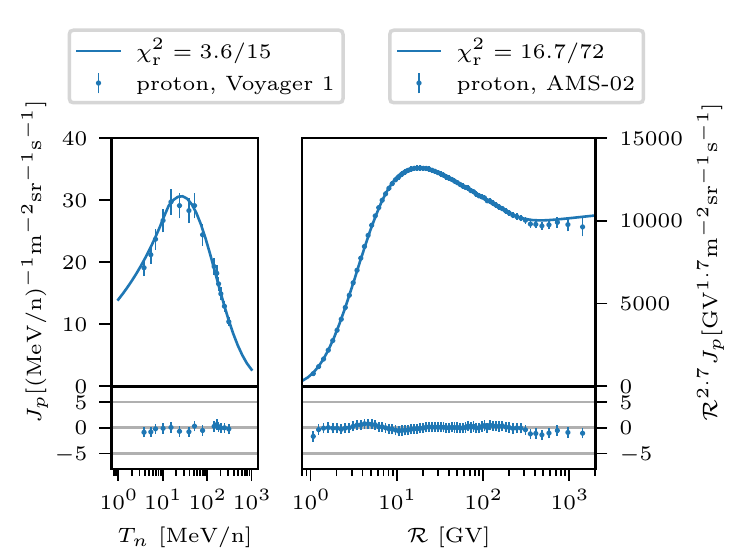} \includegraphics[scale=1]{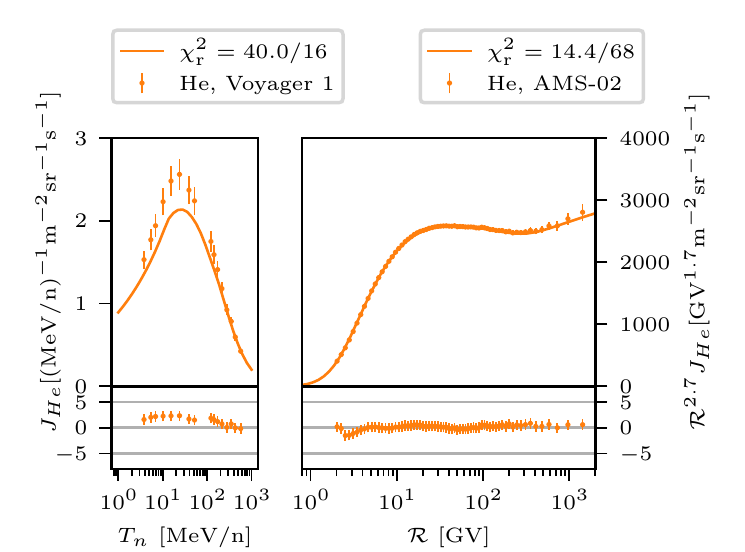} \\
\includegraphics[scale=1]{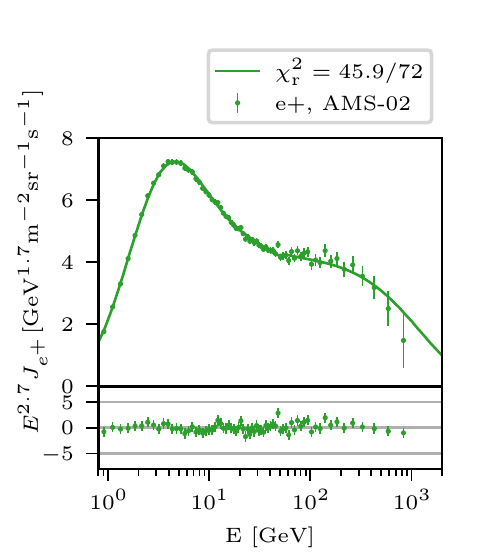} \includegraphics[scale=1]{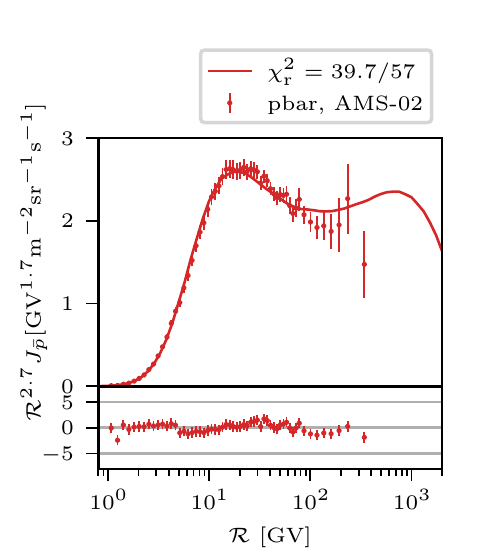} \includegraphics[scale=1]{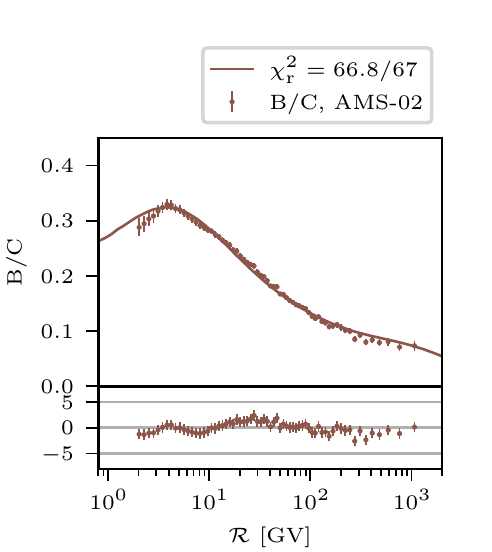} \\
\includegraphics[scale=1]{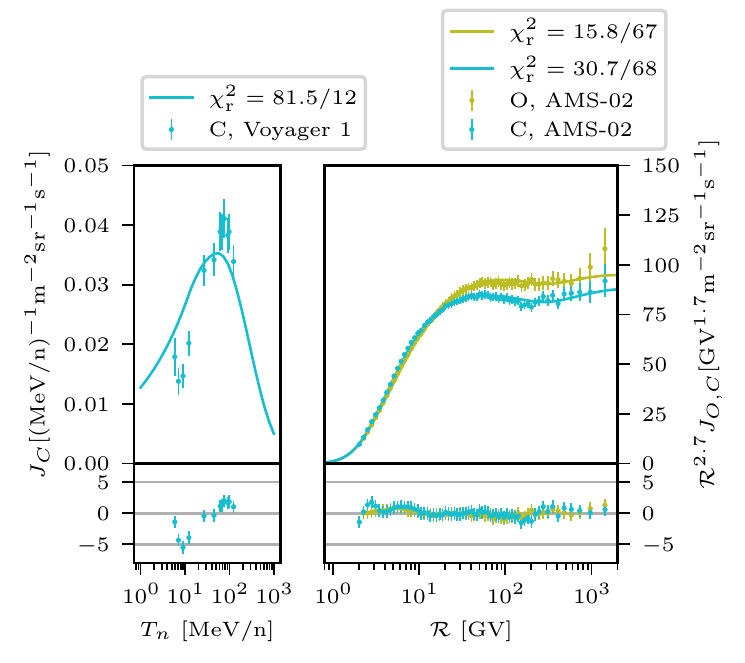} \includegraphics[scale=1]{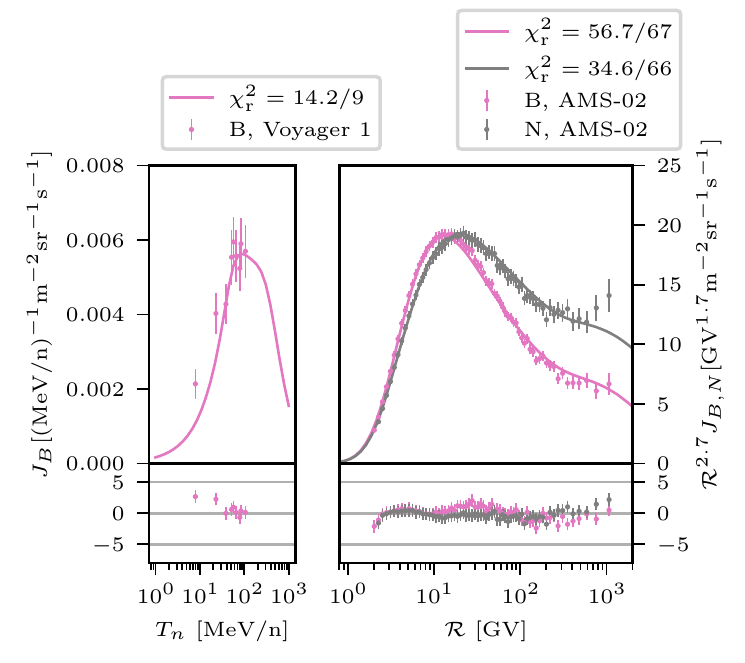}
\caption{Best-fit spectra for various CR primaries and secondaries: protons (top left), helium (top right), positrons (middle left), antiprotons (middle centre), boron-to-carbon ratio (middle right), carbon and oxygen (bottom left), boron and nitrogen (bottom right).}
\label{fig:spectra}
\end{figure*}

Our best fits are shown in Figs.~\ref{fig:spectra} and the best-fit parameters given in Tbl.~\ref{tbl1}. Overall the agreement is very satisfactory, but with a couple of notable features.

\begin{figure*}[p]
\includegraphics[scale=1,trim={0 0 11.05cm 0}, clip=true]{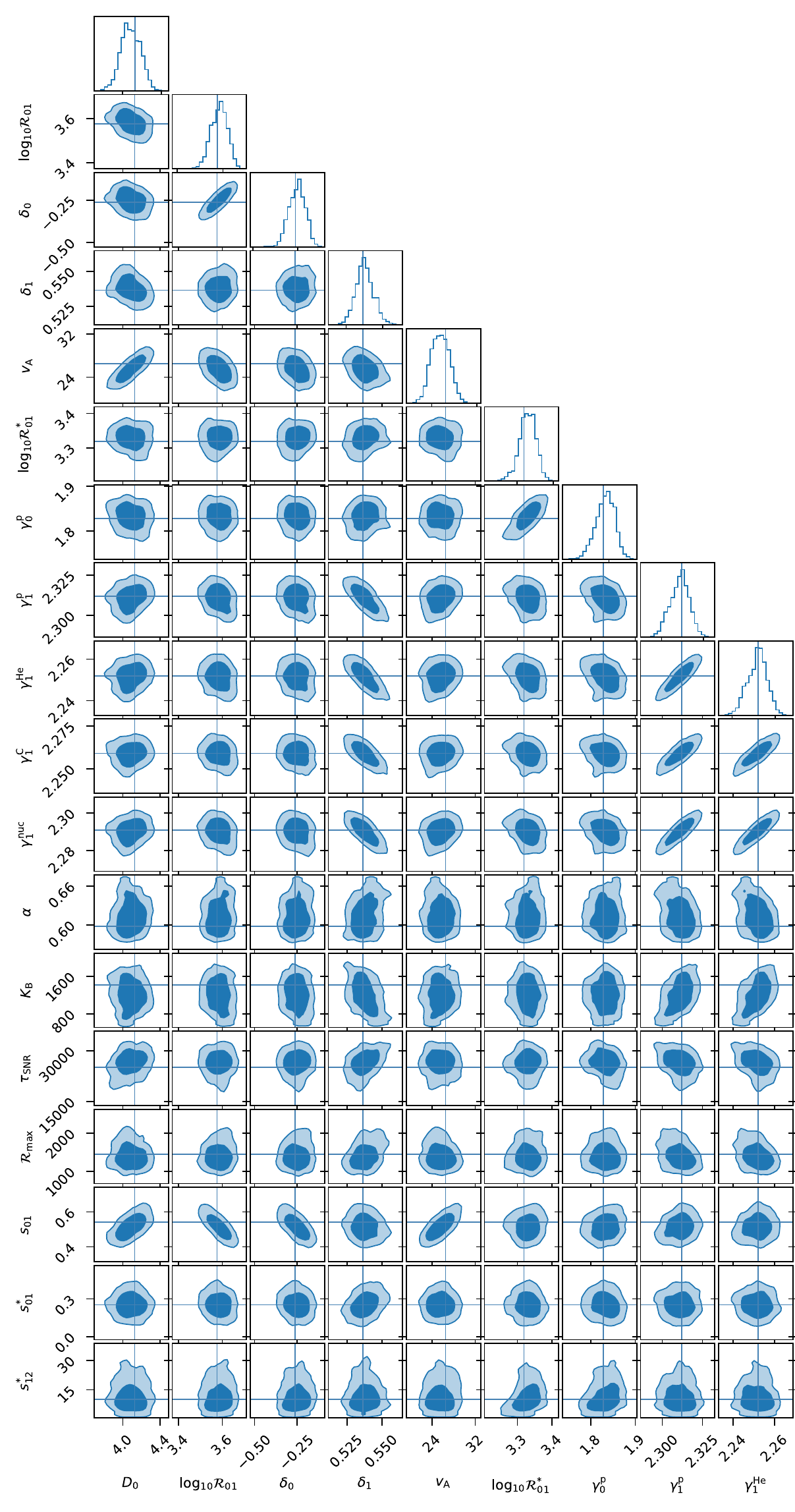}
\caption{Corner plot of the marginalised distributions of the most important free parameters of our model. Note that $D_0$ is in $10^{28} \mathrm{cm}^2 \, \mathrm{s}^{-1}$, $\mathcal{R}_{01}$ in $\mathrm{MeV}$, $v_{\mathrm{A}}$ in $\mathrm{km} \, \mathrm{s}^{-1}$, $\mathcal{R}^{*}_{01}$ in $\mathrm{MeV}$, $\tau_{\mathrm{SNR}}$ in $\mathrm{yr}$ and $\mathcal{R}_{\mathrm{max}}$ in $\mathrm{GV}$.}
\label{fig:corner1}
\end{figure*}

\begin{figure*}[thb]
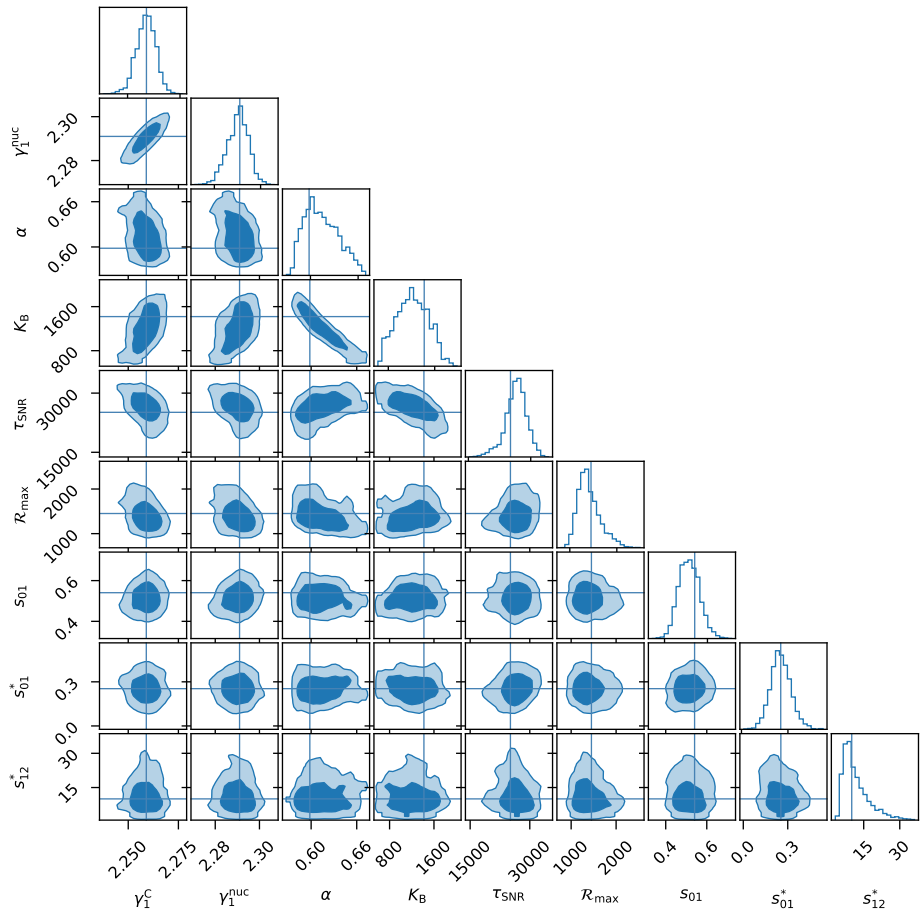

\includegraphics[scale=1,trim={0 0 21.88cm 12.35cm}, clip=true]{figures/corner-plot}
\hspace{-0.25cm}
\includegraphics[scale=1,trim={12.25cm 0 0 11.05cm}, clip=true]{figures/corner-plot}
\caption{Corner plot (cont'd from Fig.~\ref{fig:corner1}).}
\end{figure*}

The model predictions for the proton flux, both without and with solar modulation, are shown in the top left panel of Fig.~\ref{fig:spectra}. There is excellent agreement with data from both Voyager~1 (unmodulated) and AMS-02 (modulated), with a combined $\chi^2$ of 20.3 for 87 data points, even with our simplified force-field modulation model. For the AMS-02 data, adding the statistical and systematic errors in quadrature appears to overestimate the error (as is seen from the pull distribution), suggesting that some part of the systematic error is correlated in energy. 

The fit of AMS-02 helium data is equally good with a $\chi^2$ of 14.4 for 68 data points. The Voyager~1 data are matched with less fidelity at energies of a few ten MeV ($\chi^2$ of 40 for 16 data points). We stress that it is hard to improve on this, simply because the AMS-02 data points will dominate the fit even if the Voyager~1 data are included. As for the origin of this discrepancy, our emulation of the effect of source discreteness (see Sec.~\ref{sec:fit}) might be an oversimplification and we will address this issue in more detail in the future.

Our prediction for the positron flux is also an excellent match to AMS-02 data (left middle panel of Fig.~\ref{fig:spectra}, with $\chi^2$ of 45.9 for 72 data points. Again the uncertainties possibly also include some correlated systematics, but less so than for protons. 

The model also reproduces well the hard antiproton spectrum measured by AMS-02, while not overpredicting the data (middle panel of Fig.~\ref{fig:spectra}, $\chi^2$ of 39.7 for 57 data points). This is the main change with respect to our previous study \cite{Mertsch:2014poa} that was focussed on explaining the positron excess from acceleration of secondaries in old SNRs. 

To conclude we show in Fig.~\ref{fig:spectra} the fluxes of primary O and C (bottom left panel) for which the quality of the fits is comparable to those of proton and Helium. For secondary boron and nitrogen (bottom right panel) the agreement with our model prediction is also satisfactory at AMS-02 energies ($\chi^2$ of 56.7 for 67 data points, $\chi^2$ of 34.6 for 66 data points), but we caution that the relevant production cross section uncertainties are still sizeable~\cite{Genolini:2018ekk}, which might be the source of the excess visible around $T_n ~ 10$~GeV for boron. The same excess is discernible in the boron-to-carbon ratio (right middle panel of Fig.~\ref{fig:spectra}) even though the $\chi^2$ of 66.8 for 67 data points is very good.

\section{Summary and conclusions}
\label{sec:Summary}

\begin{figure*}[!thb]
\centering
\includegraphics[scale=1]{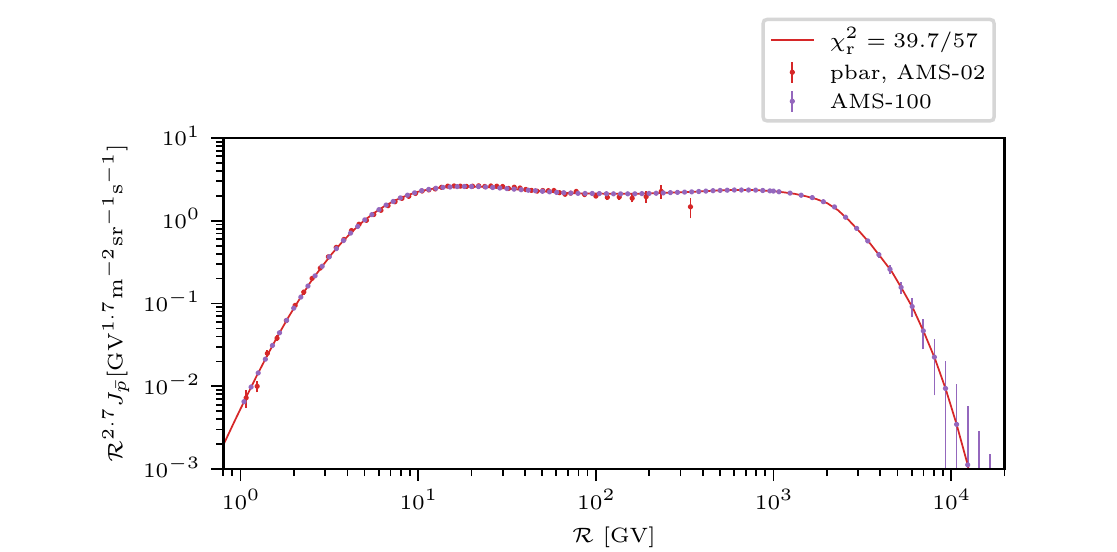}
\caption{Antiproton spectrum of Fig.~\ref{fig:spectra}, together with the data expected for the proposed AMS-100 experiment~\cite{Schael:2019lvx}.}
\label{fig:AMS100}
\end{figure*}

We have presented an updated and improved analysis of the acceleration of secondaries in old supernova remnants which was considered in our earlier work \cite{Mertsch:2009ph,Ahlers:2009ae,Mertsch:2014poa}: 

\begin{itemize}
\item The datasets that we now consider include the latest AMS-02 data on protons, helium, carbon and oxygen (all predominantly primary) as well as the secondaries boron, positrons and antiprotons. In addition, we have included data acquired by the Voyager~1 probe after it left the heliosphere in 2012, thus providing the first measurement of interstellar spectra unaffected by solar modulation.
\item We have performed a systematic study of the entire set of parameters --- both those associated with the source model, as well as those associated with propagation in the Galaxy and in the heliosphere. Earlier work had only provided fits by eye but this is no longer appropriate given the high precision of AMS-02 data. However the quality of the fits suggests that \emph{correlations} in the systematic uncertainties also need to be specified.
\item We have updated the cross sections used for the production of secondaries inside the source and in the ISM. We have investigated a number of parameterisations and have used the combination that provides the best fit to data.
\end{itemize}

Our best-fit model gives an excellent description of the AMS-02 data and is in agreement with data from Voyager~1. We conclude that consideration of the acceleration of secondaries in old SNRs can indeed account for both the observed positron excess and the unexpectedly hard antiproton spectrum. This contribution must therefore be taken into account in future fits to CR data, especially in searches for signals of exciting new physics such as dark matter annihilation or decay.

A possible way for testing this model is to identify the rather broad bump in antiprotons just below the cut-off $\mathcal{R}_{\text{max}}$. Currently, statistics are the limiting factor for a precision antiproton measurement at hundreds of GeV and beyond, but the proposed AMS-100 experiment~\cite{Schael:2019lvx} will be able to confirm or falsify this feature. AMS-100 is a large (geometrical acceptance of $\sim 100 \, \text{m}^2 \, \text{sr}$) magnetic spectrometer to be operated at Lagrange point 2, with a maximum detectable rigidity of $100 \, \text{TV}$.We show the best-fit antiproton flux together with simulated data for AMS-100 in Fig.~\ref{fig:AMS100}. It is evident that the AMS-100 experiment would be able to precisely characterise the predicted anti-proton flux, thus unambiguously clarifying the origin of the positron excess.

We stress that the agreement of our acceleration of secondaries model with recent data on secondary positrons and antiprotons warrants revisiting the expectation for antinuclei which must at some level also be produced and accelerated during DSA by SNR shocks. The only studies of this to date~\cite{Herms:2016vop,Tomassetti:2017izg} adopted much smaller values for the normalisation of the secondary production, \textit{viz.} $K_\mathrm{B} \sim 20$ as opposed to our best-fit value $K_\mathrm{B} \sim 10^3$ which is much higher. It is thus conceivable that the concommitant production of such  antinuclei may  be detectable by AMS-02, a particularly exciting prospect in view of the recent indications of events due to antihelium~\cite{Kounine:2019gsh}. We hope to return to this important issue in future work.

\section*{Acknowledgements}

We are grateful to the AMS-02 collaboration, especially Henning Gast, Andrei Kounine, Stefane Schael, Samuel Ting and Zhili Weng, for many helpful discussions and encouragement to pursue these studies.

\bibliography{accsec2020}

\end{document}